\documentclass[conference,letterpaper]{IEEEtran}

\newif\ifcomments
\commentstrue
\ifcomments
\newcommand{\rik}[1]{{\textcolor{red}{[Rik: {\em #1}]}}}
\else
\newcommand{\rik}[1]{}
\fi
\newif\ifcomments
\commentstrue
\ifcomments
\newcommand{\andrew}[1]{{\textcolor{blue}{[Andrew: {\em #1}]}}}
\else
\newcommand{\andrew}[1]{}
\fi

\newcommand{\Bin}{\mathrm{Bin}}
\newcommand{\Ham}{\mathrm{H}}
\newcommand{\lab}{\mathrm{label}}
\newcommand{\ds}{\mathrm{sig}}
\newcommand{\dsp}{\mathrm{psig}}

\newcommand{\cT}{\mathcal{T}}
\newcommand{\qh}{}

\usepackage[sort]{cite}
\usepackage{tikz}
\usetikzlibrary{decorations.pathreplacing}

\newcommand{\cS}{{\mathcal S}}
\newcommand{\cN}{{\mathcal N}}
\newcommand{\cF}{{\mathcal F}}
\newcommand{\cG}{{\mathcal G}}

\usepackage{algorithm}
\usepackage{algpseudocodex}
\usepackage[utf8]{inputenc} 
\usepackage[T1]{fontenc}
\usepackage{url}
\usepackage{ifthen}
\usepackage[cmex10]{amsmath}

\begin{document}
\title{Graph Reconstruction from \\ Noisy Random Subgraphs} 

\author{
\IEEEauthorblockN{Andrew McGregor}
\IEEEauthorblockA{University of Massachusetts Amherst\\
Amherst, MA, USA\\
Email: mcgregor@cs.umass.edu}
\and
\IEEEauthorblockN{Rik Sengupta}
\IEEEauthorblockA{IBM Research \\
Cambridge, MA, USA\\
Email: rik@ibm.com}
}

\newtheorem{definition}{Definition}[section]
\newtheorem{theorem}{Theorem}[section]
\newtheorem{lemma}{Lemma}[section]
\newtheorem{proposition}{Proposition}[section]
\newtheorem{corollary}{Corollary}[section]
\newtheorem{remark}{Remark}[section]

\maketitle


\begin{abstract}
We consider the problem of reconstructing an undirected  graph $G$ on $n$ vertices given multiple random noisy subgraphs or ``traces''. Specifically, a trace is generated by sampling each vertex with probability $p_v$, then taking the resulting induced subgraph on the sampled vertices, and then adding noise in the form of either a) deleting each edge in the subgraph with probability $1-p_e$, or b) deleting each edge with probability $f_e$ and transforming a non-edge into an edge with probability $f_e$. We show that, under mild assumptions on $p_v$, $p_e$ and $f_e$, if $G$ is selected uniformly at random, then $O(p_e^{-1} p_v^{-2} \log n)$ or $O((f_e-1/2)^{-2} p_v^{-2} \log n)$ traces suffice to reconstruct $G$ with high probability. In contrast, if $G$ is arbitrary, then $\exp(\Omega(n))$ traces are necessary even when $p_v=1, p_e=1/2$.
\end{abstract}

\newtheorem{claim}[lemma]{Claim}

\section{Introduction}\label{sec:introduction}
We consider the problem of reconstructing a graph $G$ given noisy observations of random subgraphs of $G$. We call these observations \emph{traces} and consider two different noise models: edge deletions or edge flips. Formally, we have the following.

\begin{definition}[Traces]
Given a graph $G=(V,E)$, a \emph{trace} $G'=(V', E')$ is a random graph generated as follows: first, each vertex of $G$ is sampled independently with probability $p_v$, to form $V' \subseteq V$. Then $G'$ is formed from the induced subgraph on $V'$, denoted $G[V']$, by either:
\begin{enumerate}
    \item \emph{Edge Deletion Trace}: Deleting each edge in $G[V']$ independently with probability $1-p_e$.
    \item \emph{Edge Flip Trace}: Deleting each edge in $G[V']$ independently with probability $f_e$ and adding an edge between each non-adjacent pair in $G[V']$ with probability $f_e$.
\end{enumerate}
Note that the vertices are not labeled; given two traces $G'_1$ and $G'_2$, it is impossible in general to determine whether a vertex $v\in G'_1$ and $v'\in G'_2$ correspond to the same vertex in $G$.
\end{definition}

We are interested in the number of independently generated  traces that are necessary to reconstruct the graph $G$ (with high probability). We refer to this number as the \emph{sample complexity} of reconstruction. The problem was studied by McGregor and Sengupta \cite{canon}, who considered the noiseless setting where $p_e=1$ (or equivalently $f_e=0$). They showed that $O(p_v^{-2} \log n)$ traces are sufficient for random graphs drawn from $\cG(n,1/2)$ (i.e., $n$-vertex graphs where edges are present independently with probability $1/2$), assuming $p_v=\Omega(n^{-1/6}\log^{2/3} n)$. They also showed that $2^{\Omega(n)}$ traces are necessary to distinguish arbitrary graphs even when $p_v=1/2$.

The graph reconstruction problem outlined above is partially inspired by the analogous problem for binary strings, initially proposed by Batu et al.~\cite{BatuKKM04} and subsequently studied extensively \cite{ViswanathanS08,McGregorPV14,HoldenPP18,NazarovP17,PeresZ17,holden2018lower,HolensteinMPW08,KrishnamurthyM019,DeOS17,narayanan2020circular,HartungHP18,2019arXiv190205101D,brailovskaya2021tree,maranzatto2021reconstructing}. In the case of strings, the traces correspond to random subsequences (potentially subject to further noise).
Despite extensive research, there is still a considerable gap between the best known upper and lower bounds on the sample complexity, whether the unknown string is arbitrary or random.
The other motivation for our problem is the \emph{graph reconstruction problem} from classical structural graph theory. There, the objective is to reconstruct an undirected $n$-vertex graph from the multiset of its induced subgraphs on $(n - 1)$ vertices. Determining whether this is possible for arbitrary graphs is a famous unsolved problem \cite{pjm/1103043674, Bollobs1990AlmostEG}.



\subsection{Our Results}

In this paper, we show the following upper bound on the sample complexity of reconstructing random graphs.

\begin{theorem}[Upper Bound for Random Graphs]\label{thm:mainupper}
Let $G\sim {\mathcal G}(n,1/2)$ and 
\begin{align*}
    p_v &
    = \omega(\log n/\sqrt{n}) \\
    p_e &=  \omega( p_v^{-1/3}n^{-1/6} \sqrt{\log n}) \\ 
    f_e &= 1/2-\omega( p_v^{-1/4}n^{-1/8} (\log n)^{3/8}) \ .
\end{align*} 
Then, in the edge deletion model, $4p_v^{-2} p_e^{-1} \log n$ traces are sufficient to reconstruct $G$ with probability at least $1-1/n$, where the probability is also taken over the random choice of $G$. In the edge flip model, the corresponding bound is  $12p_v^{-2} (1/2-f_e)^{-2} \log n$.
\end{theorem}

This theorem generalizes the result by McGregor and Sengupta \cite{canon}, which only applied when $p_e=1$ or $f_e=0$, i.e., when the the traces are noise-free. However, even in that setting, our approach improves upon the previous result: our algorithm is simpler, and holds for a larger range of $p_v$ values.

In Section~\ref{sec:lower}, we discuss lower bounds for reconstructing arbitrary graphs. The proof technique used in \cite{canon} to establish an $\exp(\Omega(n))$ lower bound when $p_v=1/2$ and $p_e=1$ can be modified to show that $\exp(\Omega(n))$ traces are necessary in the noisy setting where $p_e=1/2$, even when $p_v=1$. However, we conjecture that this bound can be strengthened to show that $\exp(\Omega(n^2))$ traces are necessary. We briefly discuss the challenges in proving such a result.


\section{Preliminaries}\label{sec:preliminaries}

\subsubsection{Notation and Conventions}
Let $[k]$ denote the set $\{1, \ldots, k\}$ and, for a set $S$, let $\binom{S}{k}$ denote all subsets with cardinality $k$. We only consider undirected graphs $G = (V, E)$. We use $(u,v)$ to denote an edge, and $\{u,v\}$ to denote a pair in $\binom{V}{2}$, regardless of whether $(u,v) \in E$ or not. For $v \in V$, $\Gamma_G(v)$ denotes the \emph{neighborhood} $\{v' \in V: (v, v') \in E\}$. 

Given two graphs, $G_1=(V_1,E_1)$ and $G_2=(V_2,E_2)$ where $|V_1|=|V_2|$, and a bijection $\pi: V_1\rightarrow V_2$, define the \emph{induced bijection on vertex pairs} to be $\sigma_\pi: \binom{V_1}{2} \rightarrow \binom{V_2}{2}$, where $\sigma_\pi(\{u,v\})=\{ \pi(u),\pi(v)\}$. 
Also, for $\cS\subseteq \binom{V_1}{2}$ we define:
 %
\[
\Delta_\pi^\cS:= \left |\left  \{\{u,v\}\in \cS: (u,v) \in E_1 \mbox{ iff } (\pi(u),\pi(v)) \not \in E_2 \right \} \right |\]
and $\Delta_\pi:=
\Delta_\pi^{\binom{V_1}{2}}$. The quantity $\Delta_\pi^\cS$ measures how ``far'' $\pi$ is from being an isomorphism (by mapping edges to non-edges, and vice versa). For instance, if $G_1$ and $G_2$ are isomorphic, there exists a bijection $\pi$ such that $\Delta_\pi=0$. If the mapping is clear from the context, we will suppress the subscript on $\Delta_\pi$.

Now suppose $G_1$ and $G_2$ are subgraphs of traces, and hence $V_1$ and $V_2$ are subsets of $V$. For the sake of analysis, suppose the vertices in $V$ have distinct labels, which $V_1$ and $V_2$ inherit (we reiterate that these labels are not available to our reconstruction algorithm). 

In this situation, we say $v\in V_1$ is \emph{fixed} by $\pi$ if $v\in V_1$ and $\pi(v)\in V_2$ have the same label. Otherwise, $v$ is \emph{non-fixed}. Similarly, we say a pair of vertices $\{u,v\}\in \binom{V_1}{2}$ is \emph{fixed} by $\sigma_\pi$ if $\{\lab(u),\lab(v)\}=\{\lab(\pi(u)),\lab(\pi(v))\}$. 
The following lemma\footnote{In the interest of space and readability, some of our technical proofs are relegated to the appendix.} establishes a lower bound on the number of non-fixed pairs in $\sigma_\pi$.
\begin{lemma}\label{lem:nonfixpair}
Suppose the bijection $\pi$ has $b$ non-fixed points and that $|V_1|=|V_2|=n' \geq 6$. Let $m_b$ be the the number of non-fixed pairs in $\sigma_\pi$. Then $m_b\geq b(n'-1-b/2)\geq n'b/3$.
\end{lemma}

\subsubsection{Correlated Bits and Concentration Bounds}
We will use standard notations from probability and statistics, e.g., $X \sim \Bin(N, \gamma)$ will mean the random variable $X$ is distributed according to the binomial distribution with parameters $N$ and $\gamma$. The following lemma will be used throughout the paper to quantify the probability that given two traces containing vertices $u$ and $v$, the edge $(u,v)$ is present in exactly one of them.
\begin{lemma}\label{lem:cb}
Let $X_1, X_2, Y_1, Y_2\in \{0,1\}, Z_1, Z_2, W_1, W_2\in \{-1,1\}$ be independent random variables where 
\begin{align*}
    \Pr[X_i=1] &= 1/2 \qquad \Pr[Z_i=1] =1/2 \\
    \Pr[Y_i=1] &=p_e \qquad \Pr[W_i=1] =1-f_e \ .
\end{align*}
for $i \in \{1, 2\}$.
Then,
\begin{align*}
\Pr[X_1Y_1 &\neq X_2Y_2] =p_e(1-p_e/2) \\
\Pr[X_1Y_1 &\neq X_2Y_2|X_1=X_2] =p_e(1-p_e) \\
\Pr[Z_1W_1 &\neq Z_2W_2] =1/2 \\
\Pr[Z_1W_1 &\neq Z_2W_2|Z_1=Z_2] =2f_e(1-f_e) \ .
\end{align*}
\end{lemma}

The next lemma establishes concentration bounds that we will need at multiple steps of our analysis.
\begin{lemma}\label{lem:mess}
Let $p_e\leq 1/2$ and $1/4 \leq f_e\leq 1/2$. Define: 
\begin{align*}
\gamma_1 &=p_e(1-p_e) &&
\gamma_2 =2\gamma_1/3 +\gamma_4/3 \\ 
\gamma_3 &=\gamma_1 /3+2\gamma_4/3 && 
\gamma_4 =p_e(1-p_e/2) \\
\rho_1 &=2f_e(1-f_e) &&   
\rho_2  =2\rho_1 /3  +\rho_4/3 \\
\rho_3  &=\rho_1/3 +2\rho_4/3 &&
\rho_4 ={1}/{2} \ 
\end{align*}
Then, we have:
\begin{align*}
    \Pr[\Bin(N,\gamma_1)\geq \gamma_2 N] &\leq \exp(-p_e^3 N/108) \\
    \Pr[\Bin(N,\gamma_4)\leq \gamma_3 N] &\leq \exp(-p_e^3 N/108) \\
    \Pr[\Bin(N,\rho_1)\geq \rho_2 N] &\leq \exp(-(1/2-f_e)^4 N/4) \\
    \Pr[\Bin(N,\rho_4)\leq \rho_3 N] &\leq \exp(-(1/2-f_e)^4 N/4) \ . 
\end{align*}
\end{lemma}



\subsubsection{Parameter Ranges}

In the rest of this paper, we will assume $p_e\leq 1/2$ and $f_e \geq 1/4$ to make the analysis simpler. However, our results immediately hold for larger $p_e$ and smaller $f_e$ values. This follows because, in the edge deletion model, if $p_e>1/2$, then deleting every edge in the observed traces with probability $(p_e-1/2)/p_e$ ensures that every edge is ultimately deleted with probability $(1-p_e) + p_e\cdot (p_e-1/2)/p_e=1/2$. In the edge flip model, if $f_e<1/4$ then flipping the state of every pair in a trace with probability $(1/4 - f_e)/(1 - 2f_e)$ ensures the final flip probability is 
\[
(1-f_e) \cdot \frac{1/4 - f_e}{1 - 2f_e}
+
f_e  \cdot \left  (1-\frac{1/4 - f_e}{1 - 2f_e}\right )
=1/4 \ .  
\]
We may also assume $f_e\leq 1/2$ because otherwise, we can flip the state of each pair in the traces.



\section{Reconstructing Random Graphs: \\ Edge Deletion Model} \label{sec:analysis}


To understand our approach, first suppose the vertices of the unknown graph $G$ have $n$ unique labels, and that these labels are preserved when the traces are generated. If this were the case, in the edge deletion model we would just need to ensure that we take enough traces so that every edge in the original graph would be present in at least one trace. We will shortly argue that $\Theta(p_v^{-2} p_e^{-1} \log n)$ traces are sufficient for this to hold with high probability. Unfortunately, in our setting, the vertices of the graph do not \emph{a priori} come equipped with these labels. Our main technical contribution is a systematic way to label the vertices in each trace \emph{consistently}, i.e., two vertices in different traces would receive the same label iff they correspond to the same vertex in $G$. Our approach will be to construct bijections in order to ``pair'' common vertices in each pair of traces $G_1=(V_1,E_1)$ and $G_2=(V_2,E_2)$ where $V_1,V_2\subset V$, i.e., we will be able to identify the vertices common to $V_1$ and $V_2$. Of course, if we can do this for all pairs of traces without any errors, then we can extend these bijections to equivalence classes; two vertices in different traces will be in the same equivalence class iff they correspond to the same vertex in $G$. If every vertex appears in at least one trace, then there will be exactly $n$ equivalence classes, which would give consistent labels to the vertices. Once this is done, reconstruction would be easy.

The following key lemma establishes the number of traces required to ensure that every edge in the original graph appears at least once, and shows that if we can pair vertices between each pair of traces with sufficiently high probability, then we can reconstruct the graph.

\begin{lemma}[Reconstruction via Pairing Traces]\label{pairingGDC}
Let
\begin{align*}
    p_v = \omega(\log n/\sqrt{n}) &&&&
    p_e = \omega( p_v^{-1/3}n^{-1/6} \sqrt{\log n}) \ .
\end{align*} 
Given two traces $G_1=(V_1,E_1)$ and $G_2=(V_2,E_2)$ of $G\sim {\mathcal G}(n,1/2)$, suppose that it is possible to identify the vertices in $V_1\cap V_2$ and find the correct correspondence between those vertices with probability at least $1-1/n^{10}$, where the probability is taken over the generation of $G_1$, $G_2$ and $G$. Then,
    $t:=4p_v^{-2} p_e^{-1} \log n$ traces are sufficient for reconstruction with probability at least $1-2/n^2$.
\end{lemma}
\begin{IEEEproof}
First note that \[t=o((\sqrt{n} / \log n)^{2-1/3} n^{1/6} \sqrt{\log n})\leq n \ ,\] for sufficiently large $n$, given the conditions on $p_v$ and $p_e$. By the union bound, with probability at least $1-t^2/n^{10}\geq 1-1/n^8$, we can pair up the vertices between every pair of traces.
For any $(u,v) \in E$, the probability that this edge is preserved in a given trace is $p_v^2p_e$ (since both vertices as well as the edge itself need to be preserved). So with $t$ traces, at least one of them preserves this edge with probability $1 - (1 - p_v^2p_e)^t \geq 1 - \exp(-p_v^2p_et)$. Union bounding over $n^2$ pairs gives us a probability of $1 - n^2\exp(-p_v^2p_et) = 1-n^{-2}$, since $t = 4 p_v^{-2}p_e^{-1} \log n$. So the overall success probability is $1-1/n^2-1/n^8\geq 1-2/n^2$.
\end{IEEEproof}

Algorithm \ref{alg:pair} describes our procedure for pairing two traces by matching their common vertices. Informally, given two traces $G_1 = (V_1, E_1)$ and $G_2 = (V_2, E_2)$, we find two induced subgraphs $G_1[S]$ and $G_2[T]$ with $|S| = |T| = k$, that are as close to being isomorphic as possible; specifically, 
we match their vertices in a way that minimizes the number of vertex pairs that induce an edge in one but not in the other. 
We set $k$ sufficiently large such that $k \approx |V_1\cap V_2|$. Our analysis shows that this process is guaranteed to find a large subset of the intersection $V_1\cap V_2$. 
We then augment the bijection to also match the remaining vertices in $V_1\cap V_2$. To do this, for each $v\in V_1$ and $v'\in V_2$, we generate a \emph{signature} based on $S$ and $T$ respectively, and match $v$ and $v'$ iff their  signatures are sufficiently similar. The signature is a binary vector that encodes the neighbors and non-neighbors of $v$ (resp. $v'$) amongst $S$ (resp. $T$). 
The intuition is that these vectors are sufficiently similar iff $v$ and $v'$ correspond to the same vertex in $G$.

\begin{algorithm}[!htbp]
\caption{Pairing Traces in the Edge Deletion Model}
\label{alg:pair}
\begin{algorithmic}[1]
\State Initialize $r\leftarrow \sqrt{33 p_v^2 n \log n}$. If $p_v=1$, $k\leftarrow n$ and $k\leftarrow p_v^2 n -r$ otherwise. 
 \State Given traces $G_1=(V_1,E_1), G_2=(V_2,E_2)$, find $S^\ast\subset V_1$ of size $k$, $T^\ast\subset V_2$ of size $k$, and bijection $\pi^\ast:S^\ast\rightarrow T^\ast$ that minimizes $\Delta_{\pi^\ast}$. Let $t_i = \pi^\ast(s_i)$ for all $1 \leq i \leq k$.
\State Pick an ordering of the elements in $S^\ast=\{s_1, s_2, \ldots, s_k\}$ arbitrarily. For $v\in V_1$ and $v'\in V_2$, define binary strings:
\begin{align*}
\ds_1(v) &= (I(s_1\in \Gamma_{G_1}(v)), \ldots, I(s_k\in \Gamma_{G_1}(v))) \\
\ds_2(v') &= (I(t_1\in \Gamma_{G_2}(v')), \ldots, I(t_k\in \Gamma_{G_2}(v'))) \ ,
\end{align*}
where $I(\mathcal E)$ denotes the indicator function of event $\mathcal E$.
\State 
Pair $v\in V_1$ and $v'\in V_2$ iff \[
\Ham(\ds_1(v),\ds_2(v'))\leq k (\gamma_1+\gamma_4)/2 - 1 \ , \] where $\Ham(x, y)$ is the Hamming distance between $x$ and $y$.
\end{algorithmic}
\end{algorithm}


\subsection{Analysis}

Let $G = (V, E) \sim \mathcal{G}(n, 1/2)$, $G_1 = (V_1, E_1)$, and $G_2 = (V_2, E_2)$ be defined as above.

\begin{lemma}\label{lem:A1prob}
Let $A_1$ be the event that
\begin{equation*}
    p_v^2 n - r  \leq |V_1\cap V_2|\leq p_v^2 n + r~,
\end{equation*}
where $r = \sqrt{33p_v^2 n\log n}$. Then, $\Pr[A_1]\geq 1-2/n^{11}$.
\end{lemma}

Note that if $A_1$ occurs, then $|V_1|$ and $|V_2|$ both have size at least $|V_1\cap V_2|\geq k$, and so there exists at least one triple $(S,T,\pi)$ where $S\subset V_1, T\subset V_2$, and $|S|=|T|=k$; in other words, step 2 of the algorithm returns some triple. Let $\cT$ denote the set of such triples. We next argue that with high probability the triple $(S^\ast, T^\ast, \pi^\ast)$ that minimizes $\Delta_{\pi^\ast}$ has mostly fixed points. 
To do this, we define a mapping on triples $f:\cT\rightarrow \cT$ where $f(S,T,\pi)=(S',T',\pi')$, where $S'$ is an arbitrary set of vertices satisfying:
\begin{equation*}
|S'|=k \qquad \mbox{ and } \qquad S \cap V_2  \subseteq S' \subseteq V_1\cap V_2\ .
\end{equation*}
Let $T'=S'$ and let $\pi'$ be the identity map. Note that $f$ is well-defined, as $(S', T', \pi') \in \cT$. We now show that it is very likely that $\Delta_{\pi'}$ is less than $\Delta_\pi$ if $\pi$ has many non-fixed points.

\begin{lemma}\label{lem:improvingtriple}
For any $(S,T,\pi)\in \cT$ and $(S',T',\pi') = f(S,T,\pi)$, $\Pr[\Delta_\pi >\Delta_{\pi'}] \geq 1-4\exp(-kb\cdot p_e^3/1296)$, where $b$ is the number of non-fixed points in $\pi$.
\end{lemma}
\begin{IEEEproof}
Let $\cN$ be the set of non-fixed pairs of the induced bijection $\sigma_\pi$, and let $\cF=\binom{V_1}{2}-\cN$ be the fixed pairs. Then 
$\Delta_\pi$ can be written as  $\Delta_\pi^\cN+\Delta_\pi^\cF$. Let $N=|\cN|$ and $F=|\cF|$.

\begin{claim}\label{claim:equalmatchings}
$\cN$ can be partitioned as $\cN_1\cup \cN_2\cup \cN_3$, s.t.~for all $i$, $|\cN_i| \geq N/4$ and $\Delta_\pi^{\cN_i} \sim \Bin(N_i,\gamma_4)$.
\end{claim}



It follows that $\Delta_\pi=\Delta_\pi^{\cN_1}+\Delta_\pi^{\cN_2}+\Delta_\pi^{\cN_3}+\Delta_\pi^{\cF}$. The crucial observation is that all fixed pairs in $\sigma_\pi$ are fixed pairs in $\sigma_{\pi'}$, and so $\Delta_{\pi'}=\Delta_{\pi'}^{\cF}+\Delta_{\pi'}^{\cN_1\cup \cN_2 \cup \cN_3}$ where $\Delta_{\pi'}^{\cF}=\Delta_{\pi}^{\cF}$ and $\Delta_{\pi'}^{\cN_1\cup \cN_2 \cup \cN_3} \sim \Bin(N,\gamma_4)$.
Therefore $\Pr[\Delta_\pi < \Delta_{\pi'}]$ can be bounded above as:
\begin{align*}
& \Pr\left[\sum_{i \in [3]} \Delta_\pi^{\cN_i} < \Delta_{\pi'}^{\cN_1\cup \cN_2 \cup \cN_3}\right] \\
&\leq \Pr\left[\sum_{i \in [3]} \Delta_\pi^{\cN_i}< N\gamma_3\right] + \Pr\left[\Delta_{\pi'}^{\cN_1\cup \cN_2 \cup \cN_3} > N\gamma_3\right] \\
&\leq \sum_{i \in [3]} \Pr[\Bin(N_i, \gamma_4) < N_i\gamma_3] + \Pr[\Bin(N,\gamma_1) > N\gamma_2] \\
&\leq \sum_{i \in [3]} \exp(-p_e^3 N_i/108) + \exp(-p_e^3 N/108) \ ,
\end{align*}
using Lemmas~\ref{lem:cb} and \ref{lem:mess}. This is upper bounded by:
\begin{align*}
    \Pr[\Delta_\pi < \Delta_{\pi'}] 
    &\leq 3\cdot\exp(-p_e^3 kb/1296) + \exp(-p_e^3 kb/108) \\
    &< 4\cdot\exp(-p_e^3 kb/1296)\ , 
\end{align*}
using $N_i \geq N/4$ and $N \geq kb/3$ (Lemma \ref{lem:nonfixpair}).
\end{IEEEproof}

\begin{theorem}\label{thm:A2prob}
Let $A_2$ be the event that the triple in $\cT$ minimizing $\Delta$ has no non-fixed points.
Then $\Pr[A_2|A_1]\geq 1-4n\exp(-kp_e^3/2592+2r \log n)$ assuming $4\log n \leq k \cdot p_e^3/2592$.
\end{theorem}
\begin{IEEEproof}
Let $m_b$ be the number of triples $(S,T,\pi)$ in $\cT$ where $\pi$ has $b$ non-fixed points. Let $m=|V_1\cap V_2|$. Note that there are at most $\binom{m}{k-b}n^b$ ways to pick $S$, and then given $S$, $\binom{k}{k-b} n^b$ choices for $\pi$ because we can first choose $k-b$ fixed elements of $S$, and then choose the images of the other $b$ points. This also fixes $T$. Hence, assuming $A_1$, we have:
\begin{align*}
m_b & \leq \binom{m}{k-b}n^b \binom{k}{k-b} n^b \\
& \leq \exp(2 b  \log n + b \log k +(2r+b) \log m) \\
& \leq \exp(4 b  \log n + 2r\log n) \ . 
\end{align*}
By Lemma \ref{lem:improvingtriple}, for any triple in $\cT$ with at least $b$ non-fixed points, there exists another triple with all fixed points that has a smaller value of $\Delta$ with probability at least $1-4\exp(-kbp_e^3/1296)$. So the probability there are any non-fixed points is at most $\sum_{b=1}^n 4\cdot\exp(-kbp_e^3/1296+4 b  \log n + 2r\log n)$, by the union bound.
If $4\log n \leq kp_e^3/2592$ then this is at most $4 n \exp(-k p_e^3/2592+2r \log n)$.
\end{IEEEproof}

\newcommand{\kp}{m}

\begin{theorem}\label{thm:A3prob}
Let $U=V_1\cap V_2$, $\kp=|U|$, and let $\pi_U$ be the identity map between vertices $U\subset V_1$ and $U\subset V_2$.  Pick an arbitrary ordering of $U=\{u_1, \ldots, u_{\kp}\}$. Finally, for all $v\in V_1$ and $v'\in V_2$ define:
\begin{align*}
\dsp_1(v) &= (I(u_1\in \Gamma_{G_1}(v)), \ldots, I(u_{\kp} \in \Gamma_{G_1}(v)))\ , \\
\dsp_2(v') &= (I(u_1\in \Gamma_{G_2}(v')), \ldots, I(u_{\kp} \in \Gamma_{G_2}(v')))\ .
\end{align*}

Let $A_3$ be the event that for all $v\in V_1$ and $v'\in V_2$:
\begin{align*}
v = v' &\Rightarrow
\Ham(\dsp_1(v),\dsp_2(v')) \leq \gamma_2 \kp \ , \\
v \neq  v' &\Rightarrow \Ham(\dsp_1(v),\dsp_2(v')) \geq \gamma_3 (\kp - 2) \ .
\end{align*}
Then $\Pr[A_3]\geq 1-2n^2 \exp(-p_e^3\kp/216)$.
\end{theorem}
\begin{IEEEproof}    
If $v$ and $v'$ correspond to the same vertex in $G$, then $\Ham(\dsp_1(v),\dsp_2(v'))$ is distributed as $\Bin(\kp, \gamma_1)$ or $\Bin(\kp - 1, \gamma_1)$ depending on whether or not $v\in U$. On the other hand, if $v$ and $v'$ are different vertices in $G$, then the Hamming distance is distributed as $\Bin(\kp, \gamma_4)$ (if they are both outside $U$), $\Bin(\kp - 2, \gamma_4)+X+Y$ (if they are both inside $U$), or $\Bin(\kp - 1, \gamma_4)+X$ (if one is inside $U$ and the other is outside) where  $X\sim \Bin(1,p_e/2)$ and $Y\sim \Bin(1,p_e/2)$. 

So, if $v = v'$, then using Lemmas~\ref{lem:cb} and \ref{lem:mess}, we get:
\begin{align*}
\Pr[\Ham(\dsp_1(v),\dsp_2(v')) > \gamma_2\kp] &\leq \Pr[\Bin(\kp, \gamma_1) > \gamma_2\kp] \\
&\leq \exp(-p_e^3 \kp /108) \ ,
\end{align*}
. On the other hand, if $v\neq v'$, we get:
\begin{align*}
& \Pr[\Ham(\dsp_1(v),\dsp_2(v')) < \gamma_3(\kp - 2)] \\ 
&\leq \Pr[\Bin(\kp - 2, \gamma_4) < \gamma_3(\kp - 2)] \\
&\leq \exp(-p_e^3(\kp - 2)/108) \leq \exp(-p_e^3\kp/216).
\end{align*}
Applying the union bound over $v$ and $v'$ yields the result.
\end{IEEEproof}

Recall $p_v = \omega(\log n/\sqrt{n})$ and $p_e = \omega( p_v^{-1/3}n^{-1/6} \sqrt{\log n})$. Then,
\begin{eqnarray*}
r&=&  \sqrt{33p_v^2 n\log n}= o(n p_e^3 p_v^2/\log n)\\
kp_e^3&=& p_e^3(p_v^2 n-r) = p_e^3 p_v^2 n (1-o(1))\\
n p_v^2 p_e^3&=& \omega(\log n) \ .
\end{eqnarray*}

Note that the last two of these imply that $kp_e^3 = \omega(\log n)$, and so for large enough $n$, $4\log n \leq kp_e^3/2592$, so the conditional in Theorem \ref{thm:A2prob} applies. Therefore, using Lemma \ref{lem:A1prob} and Theorems \ref{thm:A2prob} and \ref{thm:A3prob}, we have:
\begin{align*}
    & \Pr[A_1\cap A_2\cap A_3] \\
    &\geq 1-2/n^{11}  - 4n\exp(-kp_e^3/2592+2r \log n) \\ &\qquad - 2n^2 \exp(-kp_e^3/216)  \\
    & \geq 1-2/n^{11}-4n\exp(-p_v^2 n p_e^3/2592+o(p_v^2 n p_e^3))
    \\ &\qquad  - 2n^2\cdot\exp(-p_v^2 n p_e^3/216+o(p_v^2 n p_e^3)) \\
    &\geq 1-2/n^{11}-4n\exp(-\omega(\log n))
    - 2n^2\cdot\exp(-\omega(\log n)) \\
    &\geq 1-1/n^{10} \ .
\end{align*}
Assuming $A_1\cap A_2 \cap A_3$, for any $v, v'$, we have:
\begin{align*}
v=v' &\Rightarrow  \Ham(\ds_1(v), \ds_2(v')) \leq \Ham(\dsp_1(v), \dsp_2(v'))  \\
&\qquad \leq \gamma_2 m  \leq  \gamma_2 k+2r \\
v\neq v' &\Rightarrow   \Ham(\ds_1(v), \ds_2(v')) \geq \Ham(\dsp_1(v), \dsp_2(v'))-2r \\
&\qquad \geq \gamma_3 (m-2)-2r \geq \gamma_3 k-2-2r \ .
\end{align*}
Finally, note that:
\begin{equation*}
(\gamma_3-\gamma_2) k=k p_e^2/8 = \omega(\sqrt{n} p_v \log n) = \omega(r) \ ,
\end{equation*}
and so $\gamma_2 k+2r<\gamma_3 k -2r-2$ for sufficiently large $n$. Hence:
\begin{equation*}
    \frac{k(\gamma_1 + \gamma_4)}{2} - 1 = \frac{(k\gamma_2 + 2r) + (k\gamma_3 - 2r - 2)}{2} \ ,
\end{equation*}
and so the threshold in Algorithm \ref{alg:pair} always lies between $\gamma_2 k+2r$ and $\gamma_3 k -2r-2$. 

\section{Reconstructing Random Graphs:\\ Edge Flip Model}\label{sec:edgeflip}


The algorithm and analysis for the edge flip model follows along almost identical lines to those for the edge deletion model. In fact, almost all of the necessary changes are achieved by replacing every occurrence of $\gamma_i$ by $\rho_i$ and appealing to Lemmas~\ref{lem:cb} and \ref{lem:mess}. Specifically, 
\begin{enumerate}
    \item The only change in the algorithm is to replace the pairing condition to $\Ham(\ds_1(v),\ds_2(v'))\leq k(\rho_1 + \rho_4)/2 - 1$.
    \item The lower bound on the probability of $A_2$ in Theorem \ref{thm:A2prob} becomes $1-4n\exp(-kf_e(1/2 - f_e)^4/96+2r \log n)$ assuming $4\log n \leq kf_e(1/2 - f_e)^4/96$.
    \item The event $A_3$ is defined in terms of $\rho_2$ and $\rho_3$, and the lower bound for the probability of $A_3$ in Theorem \ref{thm:A3prob} becomes $1-2n^2 \exp(-\kp f_e(1/2 - f_e)^4/8)$.
\end{enumerate}
To quickly verify this, note that changing each $\gamma_i$ to $\rho_i$ and appealing to the second part of Lemma~\ref{lem:mess} results in every occurrence of $p_e^3/108$ getting replaced by $(1/2-f_e)^4/4$. With this substitution, the valid ranges for $p_v$ and $f_e$ become: 
\begin{align}
    p_v &= \omega(\log n/\sqrt{n}) \label{eq:pvrange2} \\ 
    f_e &\in [1/4, 1/2 - \omega(p_v^{-1/4}n^{-1/8}(\log n)^{3/8})] \ . \label{eq:ferange}
\end{align} 
Once these ranges are set, it is easy to verify that $k(1/2 - f_e)^4 = np_v^2(1/2 - f_e)^4(1 - o(1)) = \omega(\log n)$, so the conditional in the edge flip equivalent to Theorem \ref{thm:A2prob} applies.

Modifying the pairing proceduring (Lemma \ref{pairingGDC}) is slightly more involved. We now need enough traces so that for each pair of vertices $\{u,v\}\in \binom{V}{2}$, the majority of traces containing both nodes contain the edge $(u,v)$ iff $(u,v)\in E$.

\begin{lemma}[Reconstruction via Pairing Traces]\label{pairingBSC}
Let $p_v$ and $f_e$ satisfy Eqs.~\ref{eq:pvrange2} and~\ref{eq:ferange}. 
Given two traces $G_1=(V_1,E_1), G_2=(V_2,E_2)$, suppose that it is possible to pair the vertices in $V_1\cap V_2$ with probability at least $1-1/n^{10}$. Then, $t:=12p_v^{-2}(1/2 - f_e)^{-2}\log n$ traces are sufficient for reconstruction with probability at least $1-2/n^2$.
\end{lemma}
\begin{IEEEproof}
Note that $t=o((\sqrt{n} / \log n)^{2 - 1/2}\cdot n^{1/4}\cdot(\log n)^{-3/4}\cdot\log n) \leq n$ for sufficiently large $n$. By the union bound, with probability at least $1-t^2/n^{10} \geq 1-1/n^8$, we can pair up the nodes between every pair of traces.

Consider any $\{u,v\} \in \binom{V}{2}$. Let $X$ (resp. $Y$) be the number of traces where $u$ and $v$ are both present, and the pair $\{u, v\}$ retains (resp. changes) its state. Of course, $X \sim \Bin(t, p_v^2(1 - f_e))$, and $Y \sim \Bin(t, p_v^2f_e)$. The Chernoff bound now gives:
\begin{align*}
    \Pr[X < Y] &\leq \Pr[X \leq tp_v^2/2] + \Pr[Y \geq tp_v^2/2] \\
    &\leq \exp\left(-\left(1 - \frac{p_v^2/2}{p_v^2(1 - f_e)}\right)^2 \frac{tp_v^2(1 - f_e)}{3}\right) \\ & \qquad + \exp\left(-\left(\frac{p_v^2/2}{p_v^2f_e} - 1\right)^2 \frac{tp_v^2f_e}{3}\right) \\
    &\leq 2\exp\left(-\left(\frac{1}{2} - f_e\right)^2 \frac{tp_v^2}{3}\right) \\
    &\leq 2\exp\left(-\frac{12\log n}{3}\right) \leq 2n^{-4} \ .
\end{align*}
Taking the union bound over all $\binom{n}{2}$ pairs $\{u, v\}$ gives us a probability of $1 - \binom{n}{2}\cdot 2n^{-4} \geq 1 - 1/n^2$ of correctly identifying the state of every pair. This gives us a probability of $1 - 1/n^8 - 1/n^2 \geq 1 - 2/n^2$ for reconstruction.
\end{IEEEproof}

\section{Lower Bounds for Arbitrary Graphs}\label{sec:lower}


In this section, we consider lower bounds for reconstructing arbitrary graphs in the edge deletion model. 
For the rest of this section, assume $p_e= 1/2$ and $p_v=1$.

We first observe that the lower bound technique used in McGregor and Sengupta \cite{canon} can be modified to prove a result in this setting, thereby providing an exponential separation between the cases of random graphs and arbitrary graphs.

\begin{theorem}[Lower Bound for  Arbitrary Graphs]\label{thm:mainlower}
Consider the graphs $C_n$, an $n$-cycle, and $C_{n/2} + C_{n/2}$, the disjoint union of two $(n/2)$-cycles. Then, $\exp(\Omega(n))$ traces are necessary to distinguish them with constant probability, in the edge deletion model with $p_v=1, p_e=1/2$.
\end{theorem}



We conjecture that reconstructing arbitrary graphs actually requires $\exp(\Omega(n^2))$ traces. Note that this would match the trivial upper bound of $\exp(O(n^2))$, which is a consequence of the fact that with this many traces, one of the traces is likely to be the entire graph!


However, it seems difficult to construct two non-isomorphic graphs such that distinguishing them requires $\exp(\Omega(n^2))$ traces. For instance, consider the following plausible approach.

Let $n = 16r -8$ for a large integer $r$ and let $P_i$ denote a path graph with $i$ vertices. Let $G_1'$ be the vertex disjoint union of $r$ copies of $P_2$, $r - 1$ copies of $P_3$, $r - 1$ copies of $P_5$, and $r$ copies of $P_6$. Let $u_i$ be a leaf of the $i$th copy of $P_2$ and let $v_j$ be either of the middle vertices of the $j$th copy of $P_6$. Similarly, let $G_2'$ be the vertex disjoint union of $r - 1$ copies of $P_2$, $r$ copies of $P_3$, $r$ copies of $P_5$, and $r - 1$ copies of $P_6$. Let $w_k$ be a leaf of the $k$th copy of $P_3$ and let $x_\ell$ be the middle vertex of the $\ell$th copy of $P_5$. Let $G_1$ (resp.~$G_2$) be the complement of $G'_1$ (resp.~$G'_2$). Note that $G_1$ and $G_2$ both have $n$ vertices and are not isomorphic.

\begin{figure}[ht]
    \begin{center}
    \begin{tikzpicture}[scale=1]
 
    \tikzstyle{every node}=[draw,circle,fill=black,minimum size=5pt, inner sep=0pt, outer sep=0pt]

    \draw (0,0.5) node (b1) {};
    \draw (0.5,0.75) node (b2) [label=below:$u_i$]{};
    \draw (1,1) node (b3) [label=above:$v_j$]{};
    \draw (0.5,1.25) node (b4) {};
    \draw (0,1.5) node (b5) {};
    \draw (1.7,1) node (b6) {};
    \draw (2.4,1) node (b7) {};
    \draw (3.1,1) node (b8) {};
    \draw[thick, black] (b1) to (b2);
    \draw[thick, black] (b5) to (b3);
    \draw[thick, black] (b3) to (b8);

    \draw[very thick, red, densely dotted] (b2) to (b3);

    \draw (4,0.5) node (c1) {};
    \draw (4.5,0.75) node (c2) {};
    \draw (5,1) node (c3) [label=below:$x_\ell$]{};
    \draw (4.5,1.25) node (c4) {};
    \draw (4,1.5) node (c5) {};
    \draw (5.7,1) node (c6) [label=above:$w_k$]{};
    \draw (6.4,1) node (c7) {};
    \draw (7.1,1) node (c8) {};
    \draw[thick, black] (c1) to (c3);
    \draw[thick, black] (c5) to (c3);
    \draw[thick, black] (c6) to (c8);

    \draw[very thick, red, densely dotted] (c6) to (c3);
    		
\end{tikzpicture}
    \caption{The subgraph in $G'_1$ formed by adding $(u_i,v_j)$ is isomorphic to the subgraph formed in $G'_2$ by adding $(w_k,x_\ell)$. Therefore, in the complements $G_1$ and $G_2$, removing those edges create isomorphic graphs.}
    \label{fig:lowerboundevent}
    \end{center}
\end{figure}
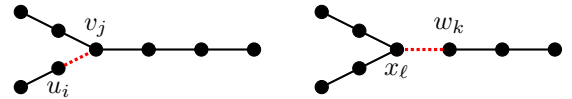

Let $E_1$ be the $r^2$ edges in $G_1$ of the form $(u_i,v_j)$, and $E_2$ be the $r^2$ edges in $G_2$ of the form $(w_k,x_\ell)$. Note that $G_1 - e_1$ is isomorphic to $G_2 - e_2$ for any $e_1\in E_1$ and $e_2\in E_2$ (see Fig.~\ref{fig:lowerboundevent}).

Note that with probability at least $1-1/2^{r^2}$ some edge $e_1$ from $E_1$ is deleted if the original graph is $G_1$ (or $e_2$ from $E_2$ if the original graph is $G_2$). Since $G_1-e_1$ is isomorphic to $G_2-e_2$, it might then seem reasonable that the variational distance between  the distributions of traces generated from $G_1$ and $G_2$ is bounded above by $2^{-O(r^{2})}$. Since $r = \Theta(n)$, it would then follow that we need $2^{\Omega(n^2)}$ traces to distinguish them. However, this turns out to not be the case.

\begin{proposition}\label{prop:arbitraryupper}
    We can distinguish between $G_1$ and $G_2$ with high probability using only $\exp(O(n^{1/3}\log^{2/3}n))$ traces in the edge deletion model with $p_v = 1$, $p_e = 1/2$.
\end{proposition}

We leave the problem of closing the gap between the upper and lower bounds as an open problem.

\section{Conclusion}\label{sec:conclusion}

Our main result was to establish an upper bound on the number of traces required to reconstruct a random graph with high probability. We  note  that our result  is optimal in the edge deletion setting, since we require $\Theta(p_v^{-2}p_e^{-1}\log n)$ traces to ensure every edge shows up at least once. It is conceivable that a similar sort of analysis can show that the theorem is also optimal in the edge flip setting.

As in several variants of the trace reconstruction problem, an important direction for future research is in the realm of time complexity. While we have optimized significantly for the sample complexity, we still require a subroutine that goes over super-exponentially many triples $(S, T, \pi)$. There may be ways of speeding up this process, but this is outside the scope of this present work.

\section*{Acknowledgment}
This work was partially supported by NSF 1934846.


\newpage
\bibliographystyle{IEEEtran}
\bibliography{bibliography}

\begin{thebibliography}{10}
\providecommand{\url}[1]{#1}
\csname url@samestyle\endcsname
\providecommand{\newblock}{\relax}
\providecommand{\bibinfo}[2]{#2}
\providecommand{\BIBentrySTDinterwordspacing}{\spaceskip=0pt\relax}
\providecommand{\BIBentryALTinterwordstretchfactor}{4}
\providecommand{\BIBentryALTinterwordspacing}{\spaceskip=\fontdimen2\font plus
\BIBentryALTinterwordstretchfactor\fontdimen3\font minus \fontdimen4\font\relax}
\providecommand{\BIBforeignlanguage}[2]{{%
\expandafter\ifx\csname l@#1\endcsname\relax
\typeout{** WARNING: IEEEtran.bst: No hyphenation pattern has been}%
\typeout{** loaded for the language `#1'. Using the pattern for}%
\typeout{** the default language instead.}%
\else
\language=\csname l@#1\endcsname
\fi
#2}}
\providecommand{\BIBdecl}{\relax}
\BIBdecl

\bibitem{canon}
\BIBentryALTinterwordspacing
A.~McGregor and R.~Sengupta, ``{Graph Reconstruction from Random Subgraphs},'' in \emph{49th International Colloquium on Automata, Languages, and Programming (ICALP 2022)}, ser. Leibniz International Proceedings in Informatics (LIPIcs), M.~Boja\'{n}czyk, E.~Merelli, and D.~P. Woodruff, Eds., vol. 229.\hskip 1em plus 0.5em minus 0.4em\relax Dagstuhl, Germany: Schloss Dagstuhl -- Leibniz-Zentrum f{\"u}r Informatik, 2022, pp. 96:1--96:18. [Online]. Available: \url{https://drops.dagstuhl.de/entities/document/10.4230/LIPIcs.ICALP.2022.96}
\BIBentrySTDinterwordspacing

\bibitem{BatuKKM04}
T.~Batu, S.~Kannan, S.~Khanna, and A.~McGregor, ``Reconstructing strings from random traces,'' in \emph{Symposium on Discrete Algorithms}, 2004.

\bibitem{ViswanathanS08}
K.~Viswanathan and R.~Swaminathan, ``Improved string reconstruction over insertion-deletion channels,'' in \emph{Symposium on Discrete Algorithms}, 2008.

\bibitem{McGregorPV14}
A.~McGregor, E.~Price, and S.~Vorotnikova, ``Trace reconstruction revisited,'' in \emph{European Symposium on Algorithms}, 2014.

\bibitem{HoldenPP18}
N.~Holden, R.~Pemantle, and Y.~Peres, ``Subpolynomial trace reconstruction for random strings and arbitrary deletion probability,'' in \emph{Conference On Learning Theory, {COLT} 2018, Stockholm, Sweden, 6-9 July 2018.}, 2018, pp. 1799--1840.

\bibitem{NazarovP17}
F.~Nazarov and Y.~Peres, ``Trace reconstruction with $\exp({O}(n^{1/3})$ samples,'' in \emph{Symposium on Theory of Computing}, 2017.

\bibitem{PeresZ17}
Y.~Peres and A.~Zhai, ``Average-case reconstruction for the deletion channel: Subpolynomially many traces suffice,'' in \emph{Symposium on Foundations of Computer Science}, 2017.

\bibitem{holden2018lower}
\BIBentryALTinterwordspacing
N.~Holden and R.~Lyons, ``{Lower bounds for trace reconstruction},'' \emph{The Annals of Applied Probability}, vol.~30, no.~2, pp. 503 -- 525, 2020. [Online]. Available: \url{https://doi.org/10.1214/19-AAP1506}
\BIBentrySTDinterwordspacing

\bibitem{HolensteinMPW08}
T.~Holenstein, M.~Mitzenmacher, R.~Panigrahy, and U.~Wieder, ``Trace reconstruction with constant deletion probability and related results,'' in \emph{Symposium on Discrete Algorithms}, 2008.

\bibitem{KrishnamurthyM019}
\BIBentryALTinterwordspacing
A.~Krishnamurthy, A.~Mazumdar, A.~McGregor, and S.~Pal, ``Trace reconstruction: Generalized and parameterized,'' \emph{{IEEE} Trans. Inf. Theory}, vol.~67, no.~6, pp. 3233--3250, 2021. [Online]. Available: \url{https://doi.org/10.1109/TIT.2021.3066010}
\BIBentrySTDinterwordspacing

\bibitem{DeOS17}
A.~De, R.~O'Donnell, and R.~A. Servedio, ``Optimal mean-based algorithms for trace reconstruction,'' in \emph{Symposium on Theory of Computing}, 2017.

\bibitem{narayanan2020circular}
\BIBentryALTinterwordspacing
S.~Narayanan and M.~Ren, ``{Circular Trace Reconstruction},'' in \emph{12th Innovations in Theoretical Computer Science Conference (ITCS 2021)}, ser. Leibniz International Proceedings in Informatics (LIPIcs), J.~R. Lee, Ed., vol. 185.\hskip 1em plus 0.5em minus 0.4em\relax Dagstuhl, Germany: Schloss Dagstuhl--Leibniz-Zentrum f{\"u}r Informatik, 2021, pp. 18:1--18:18. [Online]. Available: \url{https://drops.dagstuhl.de/opus/volltexte/2021/13557}
\BIBentrySTDinterwordspacing

\bibitem{HartungHP18}
L.~Hartung, N.~Holden, and Y.~Peres, ``Trace reconstruction with varying deletion probabilities,'' in \emph{Workshop on Analytic Algorithmics and Combinatorics}, 2018.

\bibitem{2019arXiv190205101D}
\BIBentryALTinterwordspacing
S.~Davies, M.~Z. Racz, and C.~Rashtchian, ``Reconstructing trees from traces,'' in \emph{Proceedings of the Thirty-Second Conference on Learning Theory}, ser. Proceedings of Machine Learning Research, A.~Beygelzimer and D.~Hsu, Eds., vol.~99.\hskip 1em plus 0.5em minus 0.4em\relax Phoenix, USA: PMLR, 25--28 Jun 2019, pp. 961--978. [Online]. Available: \url{http://proceedings.mlr.press/v99/davies19a.html}
\BIBentrySTDinterwordspacing

\bibitem{brailovskaya2021tree}
\BIBentryALTinterwordspacing
T.~Brailovskaya and M.~Z. R{\'{a}}cz, ``Tree trace reconstruction using subtraces,'' \emph{CoRR}, vol. abs/2102.01541, 2021. [Online]. Available: \url{https://arxiv.org/abs/2102.01541}
\BIBentrySTDinterwordspacing

\bibitem{maranzatto2021reconstructing}
\BIBentryALTinterwordspacing
T.~Maranzatto and L.~Reyzin, ``Reconstructing arbitrary trees from traces in the tree edit distance model,'' \emph{CoRR}, vol. abs/2102.03173, 2021. [Online]. Available: \url{https://arxiv.org/abs/2102.03173}
\BIBentrySTDinterwordspacing

\bibitem{pjm/1103043674}
\BIBentryALTinterwordspacing
P.~J. Kelly, ``A congruence theorem for trees.'' \emph{Pacific Journal of Mathematics}, vol.~7, pp. 961--968, 1957. [Online]. Available: \url{https://api.semanticscholar.org/CorpusID:55091877}
\BIBentrySTDinterwordspacing

\bibitem{Bollobs1990AlmostEG}
\BIBentryALTinterwordspacing
B.~Bollob{\'a}s, ``Almost every graph has reconstruction number three,'' \emph{J. Graph Theory}, vol.~14, pp. 1--4, 1990. [Online]. Available: \url{https://api.semanticscholar.org/CorpusID:43506446}
\BIBentrySTDinterwordspacing

\end{thebibliography}

\newpage

\appendices

 \section{Omitted Proofs from Section \ref{sec:preliminaries}}

\begin{IEEEproof}[Proof of Lemma~\ref{lem:nonfixpair}]
All pairs in $\binom{V_1}{2}$ involving exactly one vertex not fixed by $\pi$ are also non-fixed in $\sigma_\pi$. There are exactly $b(n'-b)$ such pairs. There are $\binom{b}{2}$ pairs in $\binom{V_1}{2}$ where neither vertex is fixed by $\pi$. There can be at most $b/2$ of these pairs that are fixed in $\sigma_\pi$ (by the vertices mapping to each other). All the other pairs must form non-fixed pairs. It follows that:
\begin{equation*}
    m_b \geq \binom{b}{2} - \frac{b}{2} + b(n'-b) = b(n'-1-b/2) \geq bn'/3.\qh
\end{equation*}
\end{IEEEproof}

\begin{IEEEproof}[Proof of Lemma~\ref{lem:cb}]
We have:
\begin{align*}
    &\Pr[X_1Y_1\neq X_2Y_2] \\
    &\qquad= 1 - \Pr[X_1Y_1 = X_2Y_2 = 1] - \Pr[X_1Y_1 = X_2Y_2 = 0] \\
    &\qquad= 1 - p_e/2\cdot p_e/2 - (1 - p_e/2)^2 \\
    &\qquad= p_e(1 - p_e/2) \ .
\end{align*}
Similarly, we have:
\begin{align*}
    &\Pr[X_1Y_1\neq X_2Y_2 | X_1 = X_2] \\
    &\qquad= \Pr[X_1 = X_2 = 1 \land Y_1 \neq Y_2]/\Pr[X_1 = X_2] \\
    &\qquad= (1/2)^2(2p_e(1 - p_e))/(1/2) \\
    &\qquad= p_e(1 - p_e) \ .
\end{align*}
To prove the last two equations note that \[\Pr[Z_1W_1\neq Z_2W_2]=\Pr[Z_1/Z_2 \neq W_2/W_1]=1/2 \ ,\] since after fixing $W_2/W_1=w\in\{-1,1\}$, $Z_1/Z_2$ is uniform in $\{-1,1\}$ and hence equals $w$ with probability $1/2$.
Lastly, 
\[
\Pr[Z_1W_1\neq Z_2W_2|Z_1=Z_2]
\Pr[W_1\neq Z_2]=2f_e(1-f_e) \ .
\]
\end{IEEEproof}

\begin{IEEEproof}[Proof of Lemma~\ref{lem:mess}]
For $p_e\leq 1/2$ we have $0\leq \gamma_2/\gamma_1-1\leq 1$. By an application of the Chernoff bound, we get:
\begin{align*}
& \Pr[\Bin(N,\gamma_1)\geq \gamma_2 N] \\
& = \Pr\left[\Bin(N,\gamma_1)-\gamma_1N \geq \left(\frac{\gamma_2}{\gamma_1}-1\right) N\gamma_1\right] \\
& \leq 
\exp\left(-\left(\frac{\gamma_2}{\gamma_1}-1\right)^2\cdot N\gamma_1/3\right) \\
& \leq \exp(-(\gamma_2-\gamma_1)^2 N/(3\gamma_1)) \ .
\end{align*}
Then applying $\gamma_2-\gamma_1=(\gamma_4-\gamma_1)/3=p_e^2/6$ and $\gamma_1\leq p_e$ gives the result. Similarly, we have:
\begin{align*}
\Pr[\Bin(N,\gamma_4)\leq \gamma_3N]
& \leq \exp(-(\gamma_4-\gamma_3)^2  N/(3\gamma_4))  \\ & \leq \exp (-p_e^3 N/108) \ .
\end{align*}
For $1/4 \leq f_e\leq 1/2$ we have $0\leq \rho_2/\rho_1-1\leq 1$.  By applying the Chernoff bound, we get:
\[
\Pr[\Bin(N,\rho_1)\geq \rho_2 N]
\leq \exp(-(\rho_2-\rho_1)^2 N/(3\rho_1))
\]
\[
\Pr[\Bin(N,\rho_4)\leq \rho_3 N]
\leq \exp(-(\rho_4-\rho_3)^2 N/(3\rho_4))
\]
and substituting $\rho_4-\rho_3=\rho_2-\rho_1=(\rho_4-\rho_1)/3=2(1/2-f_e)^2/3$ and $\rho_1\leq \rho_4=1/2$ gives the claimed bounds.
\end{IEEEproof}

\section{Omitted Proofs from Section \ref{sec:analysis}}

\begin{IEEEproof}[Proof of Lemma \ref{lem:A1prob}]
Let $X=|V_1\cap V_2|$ and note that $X \sim \Bin(n,p_v^2)$. Then, by Chernoff, we have:
 \begin{align*}
&  \Pr[|X-p_v^2 n| \geq r] \\
 &= \Pr[|X-p_v^2 n| \geq \sqrt{33 \log (n)/(p_v^2 n)} \cdot p_v^2 n ] \\
 &\leq 2\cdot\exp\left( -33\cdot\frac{\log n}{p_v^2 n} \cdot\frac{p_v^2 n}{3}\right) = 2/n^{11} \ .\qh 
\end{align*}
\end{IEEEproof}

\begin{IEEEproof}[Proof of Claim \ref{claim:equalmatchings}]
Construct a directed graph $G_\cN$ as follows:
\begin{itemize}
    \item the nodes of $G_\cN$ are the pairs $\{u, v\} \in \cN$;
    \item there is an arc in $G_\cN$ from the pair $\{u, v\}$ to the pair $\{u', v'\}$ if and only if $\sigma_\pi(\{u, v\}) = \{u', v'\}$.
\end{itemize}
Each node in $G_\cN$ contributes $0$ or $1$ to the flipped pair count. Observe that if two nodes in $G_\cN$ are not the endpoints of an arc, then their contributions to the flipped pair count are independent of each other, as they correspond to two distinct pairs in $\cN$, which map to two other pairs (in $\binom{V}{2}$ but not necessarily in $\cN$).
Therefore, an independent set in $G_\cN$ corresponds to pairs in $\cN$ that contribute $0$ or $1$ to $\Delta_\pi$ independently of each other. What is the probability of such a pair contributing $1$ to $\Delta_\pi$? Either the pair was an edge (probability $p_e/2$) that goes to a non-edge (probability $(1 - p_e/2)$), or it was a non-edge (probability $(1 - p_e/2)$) that goes to an edge (probability $p_e/2$), for a total probability of $p_e(1 - p_e/2) = \gamma_4$. Therefore, the contribution from any independent set of size $\ell$ in $G_\cN$ is distributed as $\Bin(\ell, \gamma_4)$.

Therefore, it suffices to prove that $V(G_\cN)$ can be partitioned into three independent sets $\mathcal{I}_1 \cup \mathcal{I}_2 \cup \mathcal{I}_3$ such that $\max_i|\mathcal{I}_i| - \min_i|\mathcal{I}_i| \leq 1$.

We claim first that every node in $G_\cN$ has in-degree at most $1$ and out-degree at most $1$. Otherwise, some node corresponds to a pair in $\cN$ that appears in the domain or image of $\sigma_\pi$ twice, contradicting the fact that $\sigma_\pi$ is a bijection. Therefore, $G_\cN$ is a disjoint union of directed paths and directed cycles (including, possibly, $2$-cycles).

We now algorithmically partition $V(G_\cN)$ into three sets $\mathcal{I}_1 \cup \mathcal{I}_2 \cup \mathcal{I}_3$, maintaining the invariant throughout that
\begin{itemize}
    \item $\mathcal{I}_1$, $\mathcal{I}_2$, and $\mathcal{I}_3$ are all independent sets, with $|\mathcal{I}_1| \geq |\mathcal{I}_2| \geq |\mathcal{I}_3| \geq |\mathcal{I}_1| - 1$.
\end{itemize}
First initialize $\mathcal{I}_1 = \mathcal{I}_2 = \mathcal{I}_3 = \emptyset$ (note that the invariant is met), and take the components of $G_\cN$ in some order.
\begin{itemize}
    \item If this component is a directed path, sweep the path from one end, putting the vertices in $\mathcal{I}_3$, $\mathcal{I}_2$, and $\mathcal{I}_1$ in that (cyclic) order.
    \item If this component is a cycle with length $\equiv 0\pmod 3$, start with any vertex of the cycle, and put them in $\mathcal{I}_3$, $\mathcal{I}_2$, and $\mathcal{I}_1$ in that cyclic order.
    \item If this component is a cycle with length $\equiv 1\pmod 3$, put any node of the cycle in $\mathcal{I}_3$. Now start with its out-neighbor and put the rest into $\mathcal{I}_2$, $\mathcal{I}_3$, and $\mathcal{I}_1$ in that cyclic order.
    \item If this component is a cycle with length $\equiv 2\pmod 3$, put any node $u$ of the cycle in $\mathcal{I}_2$, and put its out-neighbor $v$ in $\mathcal{I}_3$. Now start with the out-neighbor of $v$, and put the rest into $\mathcal{I}_2$, $\mathcal{I}_3$, and $\mathcal{I}_1$ in that cyclic order.
\end{itemize}
After each step, we renumber the sets so that $|\mathcal{I}_1| \geq |\mathcal{I}_2| \geq |\mathcal{I}_3| \geq |\mathcal{I}_1| - 1$. It is now easy to see that the invariant is met at the end of each step. Therefore, this process terminates with the desired partition.
\end{IEEEproof}

\section{Omitted Proofs from Section \ref{sec:lower}}
\begin{IEEEproof}[Proof of Theorem \ref{thm:mainlower}]
Suppose that the vertices of $G_1=C_n$ are numbered $u_1, \ldots, u_n$ in order, and suppose the vertices of $G_2=C_{n/2} + C_{n/2}$ are labeled $v_1, \ldots, v_{n/2}$ in one cycle, and $v'_1, \ldots, v'_{n/2}$ in the other cycle. Let $D_1$ (resp.~$D_2$) be the distribution over traces from $C_{n}$ (resp.~$C_{n/2} + C_{n/2}$). 
Consider the following coupling argument applied to $G_1$ and $G_2$ simultaneously. For $i=1$ to $n/2$, set $d_i=1$ with  probability $1/4$ and 0 otherwise and  if $d_i=1$ delete the following four edges: \[\{(u_i, u_{i+1}), (u_{n+i},u_{n+i+1}), (v_i,v_{i+1}),  (v_i',v'_{i+1})  \}. \] Then, for $i=1$ to $n/2$, if $d_i=0$ then with probability 2/3, randomly delete either $\{(u_i, u_{i+1}), (v_i,v_{i+1})\}$ or  $\{(u_{n+i}, u_{n+i+1}), (v_i',v'_{i+1})\}$.
Note that the trace of $G_1$ and $G_2$ generated in this way is distributed according to $D_1$ and $D_2$ respectively. Note that if there exists $i$ such that $d_i=1$ then the two traces generated by this coupled process are equal. Since the probability that there exists such an $i$ is $1-(3/4)^{n/2}$, it follows that the variational distance between $D_1$ and $D_2$ is $\exp(-O(n))$.
It follows (see, e.g., \cite[Lemma A.5]{holden2018lower}) that we need at least $\exp(\Omega(n))$ traces to distinguish between $D_1$ and $D_2$ with constant probability.
\end{IEEEproof}



\begin{IEEEproof}[Proof of Proposition \ref{prop:arbitraryupper}]
For a subset of vertices $S$, define $\textup{cut}(S)$ to be the number of edges with exactly one endpoint in $S$. Define the $t$-cut distribution to be the multiset $\{\textup{cut}(S): |S|=t\}$. For $t=2$, let $c_1, c_2, \ldots, c_{\binom{n}{2}}$ be the entries of this multiset. Consider the random process of generating a trace (recall that we are assuming that $p_v=1$ and $p_e=1/2$), randomly selecting two vertices $S=\{u,v\}$ in the trace, and returning the number of edges in the trace that have exactly one endpoint in $S$. Call this number $\tilde{c}(S)$ and note that $\tilde{c}(S)\sim \Bin(c(S),1/2)$. Since $S$ is equally likely to be any of the $\binom{n}{2}$ pairs of vertices, this process is equivalent to drawing a value $c_i$ from the multi-set $\{c_1, c_2, \ldots, c_{\binom{n}{2}}\}$, and then return a value drawn from  $\Bin(c_i,1/2)$. As such, the random output is a draw from a mixture of binomials and we can use an algorithm  by Krishnamurthy et al.~\cite{KrishnamurthyM019} to learn the 2-cut distribution using $\exp(O(n^{1/3} \log^{2/3}))$ traces. Note that this would allow us to distinguish $G_1$ and $G_2$ since these graphs have different $2$-cut distributions. Specifically, the number of sets $S\subset \binom{V}{2}$ such that $c(S)=2(n-2)$ is $r$ in the case of $G_1$ and $r-1$ in the case of $G_2$. Hence, it is possible to distinguish $G_1$ and $G_2$ with $\exp(O(n^{1/3} \log^{2/3}n))$ traces.
\end{IEEEproof}

\end{document}